\documentclass[aps,twocolumn,groupedaddress]{revtex4}
\usepackage{amsmath,amssymb}

\begin{document}
\bibliographystyle{apsrev}


\title{Flashing annihilation term of a logistic kinetic
as a mechanism leading to Pareto distributions.}




\author{Ryszard Zygad\l{}o }
\affiliation{Marian Smoluchowski
Institute of Physics, Jagiellonian University, Reymonta 4,
 PL--30059 Krak\'ow,
Poland}


\date{\today}

\begin{abstract}
It is shown analytically that the flashing annihilation
term of a Verhulst kinetic leads to the power--law
distribution in the stationary state. For the
frequency of switching slower than twice the free growth rate
this provides the quasideterministic source of a Levy
noises at the macroscopic level.
\end{abstract}

\maketitle


\section{Introduction}
$1/f$ noise [1--6] and power--law distributions [6--9] are widely
 observed
signatures of {\it anomalous} behavior [7,10]. Such phenomena are
related to the scale invariance of considered systems and appear in
different domains of natural and social sciences.
Their universal character is partially explained on the statistical
 manner as
a consequence of the central limit theorem for scaleless processes,
where the Gaussian distribution is replaced by appropriate
 $\alpha$-stable
distribution [7,10]. This corresponds to the use of the
Levy noises to describe fluctuations or fractional dynamics at the
 level
of Fokker-Planck equation [11,12]. Anomalous behavior can also be
 obtained
for nonlinear systems driven by (conventional)
multiplicative Gaussian white noise [8,6], particularly for
Lotka--Volterra (or Malthus--Verhulst) model [8]. In any case
processes exhibiting fat--tailed distributions are subjected to
 large fluctuation,
so within certain stochastic description it is difficult to
distinguish to which extent such behavior is an intrinsic property
of the kinetics and to which extent it is simply the reflection of
 the presence
(and specific properties) of noise. The source of the
possibly intense noise is known for thermodynamical critical
 systems.
The search for explanation of scale invariant phenomena in {\it
 generic}
systems is still unfinished [2--6,8].
Particularly, it is suggested in Ref.~[2], see also Ref.~[5], that
 the $1/f$
noise in a membrane channel is produced by random switching between
both conducting states, rather than by inherent properties of ion
 transport.
Similarly, the importance of activation--deactivation processes is
 addressed
in Ref.~[3].

In the present paper we are going to show analytically that
the asynchronic switching between generic pure Malthusian
and Verhulst--type
kinetics leads to the Pareto distribution in the stationary state.
The result seems interesting
for few reasons. The logistic equation is used, which is
the basic model of evolution for social sciences.
The power--law distribution was identified by Pareto to describe
 the
social statistics, namely the distribution of (large) wealths.
The model is {\it quasideterministic}, so the asymptotic
power--law behavior has the kinetic origin only, coming as a
 result of
balance between Malthusian growth and Verhulstian saturation.
Finally, our result supports the general opinion of Ref.~[2] about
possible sources of anomalous behavior.

\section{Model}
Power--law behavior may be considered as an intermediate one
between divergent and bounded kinetics. Let us note that
the Malthus equation,
\begin{equation}            \dot{x}_t= ax,
  \end{equation}
where $a$ is a positive difference between
birth and death rate, leads to the exponential growth of
 population.
Taking into account the Verhulst competition coefficient $bx^\mu$,
$\mu>0$ (and equal unity for the ``true'' Verhulst model),
depending on the actual size of population,
\begin{equation}            \dot{x}_t= ax-b x^{\mu +1},
  \end{equation}
one obtains, instead of unbounded trajectories, the monotonic
 relaxation
to the stationary value
\begin{equation}
x_{{\rm st}}= (a/b)^\nu, \qquad  \nu=1/\mu,
\end{equation}
with the relaxation time
\begin{equation}
T=1/\mu a.
\end{equation}
So, if the annihilation term of Eq.~(2) is absent ($b=0$)
the kinetic is divergent and as long as $b>0$ the kinetic
is bounded. Thus one expects that the ``flashing''
annihilation term, temporarily switched on and off, may
result in a power-law distribution
for the asymptotic state. Let us simply assume a
(Markovian) binary and asynchronic character of a
switching process
\begin{equation}
b(t) = b[1+\sigma I_t],
\end{equation}
where $I_t = (-1)^{{\cal
 N}_t}$, where
${\cal N}_t$ is a Poisson process with parameter $\lambda$
and $\sigma =
 +1$ or $-1$ with probability $1/2$. It means that
the evolution consist
of periods of active annihilation (with a coefficient $2b$)
separated by periods of a free growth. The length of
the periods is random with the average equal $\lambda^{-1}$.

\section{Critical $\sigma^2$--dependence of moments}

The Bernoulli equation (2) driven by dichotomous Markov process has
been already considered [13,14]. Particularly,
in Ref.~[14] the
transient behavior of (2) with nonlinear coupling (5), $|\sigma|\le 1$,
has been examined and
in the limit $t \to \infty$
the following formula for the stationary moments
\begin{equation}
\langle x^s \rangle_{{\rm st}} = x_{{\rm st}}^s\,  {}_2 F_1 \Bigl(
 {s\nu\over
2};
{s\nu+1\over2};
1+\delta;
{\sigma}^2 \Bigr)\,
\end{equation}
where
\begin{equation}
\delta = \lambda/\mu a -1/2,
\end{equation}
was obtained.

The value $\sigma^2 =1$ is
critical with respect to the convergence of the
hypergeometric series (6)
({\it branch point} of the function)
and for the critical exponent
\begin{equation}
1+\delta -s\nu -1/2 \equiv (\lambda/ a -s)\nu \le 0,
\end{equation}
the r.h.s. of Eq.~(6) diverges at $\sigma^2=1$.
The condition (8) clearly shows the existence
of fat--tail
\begin{equation}
P(x) \sim x^{-1-\lambda/ a}
\end{equation}
for large $x$.

\section{Stationary probability density}
The stationary probability density is formally given by the inverse
Mellin transformation (with respect to $s$) of Eq.~(6).
The more convenient way of computation is to
apply the so--called quadratic transformation of
the hypergeometric function [15]
\begin{equation}
\langle x^s \rangle_{{\rm st}} = x_{{\rm st}}^s (1+
 \sigma)^{-\nu s} \,
{}_2 F_1
\Bigl(
s\nu;
{1\over2}+\delta;
1+2\delta;
{2\sigma \over 1+\sigma}   \Bigr)\,
\end{equation}
and then use of Euler's integral representation
\begin{equation}
\langle x^s \rangle_{{\rm st}} = {x_{{\rm st}}^s  (1+
 \sigma)^{-\nu s}
\over B(\gamma,\gamma)} \int\limits_0^1 dt t^{\gamma-1}
 (1-t)^{\gamma-1}
\Bigl[1- {2\sigma t \over 1+\sigma}\Bigr]^{-\nu s} ,
\end{equation}
where $\gamma= 1/2+ \delta$ and $B$ is the Beta function.
 Introducing a new
variable
$\xi
=
[b(1+\sigma
-
2\sigma t)/a]^{-\nu}$  we obtain finally
\begin{equation}
\langle x^s \rangle_{{\rm st}} = \int_{x_l}^{x_r} d\xi P
 (\xi) \xi^s
\,,
\end{equation}
where
\begin{equation}
P(x) = N^{-1} x^{-\mu -1} \bigl[ \sigma^2 - (1-
 ax^{-\mu}/b)^2
\bigr]^{\delta -1/2},
\end{equation}
and $x \in (x_l, x_r)$,
\begin{equation}
x_l = \Bigl[ {a \over b(1+|\sigma|)}\Bigr]^\nu, \qquad
x_r = \Bigl[ {a \over b(1-|\sigma|)}\Bigr]^\nu.
\end{equation}
The normalization constant in  Eq.~(13) is equal $N = B(\gamma,
 \gamma)
(2|\sigma|)^{2\delta} b/(\mu a)$. Note that in the case of
interest, $|\sigma| =1$, the right boundary $x_r \to \infty$ and
\begin{equation}
 P(x) \propto x^{-1- \lambda/a} (2- ax^{-\mu}/b)^{\delta -1/2}\,,
\end{equation}
where $x \in \bigl( (a/2b)^\nu,\, \infty \bigr)$, in agreement
with Eq.~(9).

\section{Noise--induced transitions}
Let us remind that in the present work we consider
Eqs.~(2), (5) with $|\sigma|= 1$ as a noiseless evolution
consisting
of two essentially different
(and randomly switched) modes of deterministic kinetics.
On the other hand the same model,
especially for $|\sigma| \ll 1$, may be treated as a stochastic kinetic
with nonlinearly coupled parametric noise, when the fluctuation of a
coefficient $b$ are described by a zero mean dichotomous color noise
$\xi_t \equiv b\sigma I_t$, $\langle \xi_t \xi_0 \rangle =
b^2 \sigma^2 e^{-2\lambda t}$, of a correlation--time $\tau$
and intensity $D$,
\begin{equation}
\tau=(2\lambda)^{-1}, \qquad D= b^2 \sigma^2 \tau.
\end{equation}
From this point of view it
is worth to analyze the $\tau$ and $D$ dependence
within the context of
so--called noise--induced phase transitions [13].
The case $\sigma^2 =1$ correspond to the
critical line
$b^{-2} D/\tau =1$ in Fig.~1 and to
the ``true'' transition, which is reflected
by {\it singular} $\sigma^2$--dependence
of observables, Eq.~(6), and related power--laws.
Below this line the kinetic is bounded and
consequently the
support of stationary distribution is finite, Eq.~(14).
Above the line the system is {\it unstable} and
$x_t$ rapidly approaches infinity (after finite time).
At the critical line the stationary states are
the semiaxis distributions (lower curves in Fig.~1),
Eq.~(15),
exhibiting fat--tails with indices $\alpha= \lambda/a$.
The dependence on the correlation--time is also interesting.
If the frequency of switching $\lambda$ is small
(long correlations),
$
\delta -1/2 \equiv \lambda/\mu a -1 < 0,
$
both boundaries $x_l$ and $x_r$ --- which are the stationary values (3)
of a deterministic kinetics (2) with coefficients $b(1+|\sigma|)$
and $b(1-|\sigma|)$ respectively ---
are attractive,  the probability density (13) has a minimum and it is
convex down.
In contrast for $\lambda$ greater than the
relaxation rate $\mu a$, Eq.~(4), the boundaries become repulsive.
$P(x)$ is then unimodal. The $\lambda =\mu a$ [or $\tau=1/(2\mu a)$]
is a critical value with
respect to the noise--induced transition in probability density.

\section{Exact Pareto distribution}
The case $\lambda=\mu a$, see
the vertical line in Fig.~1,
is exceptional for other reasons.
The stationary distribution is then purely power,
$P(x) \propto x^{-1-\mu}$, and neither diverges nor vanishes at
the boundaries (14).
And for $|\sigma|=1$, at the {\it critical point} common for both
lines, it becomes
the exact Pareto distribution
\begin{equation}
P(x) = (a\mu/2b) x^{-1-\mu}, \qquad x \ge (a/2b)^\nu.
\end{equation}
Moreover the case $\lambda=\mu a$
was the nontrivial one
for which the time-dependent transient moments were
found in the closed analytical
form [14]
\begin{eqnarray}
&&\langle x_t^s \rangle  =
e^{-\mu at} [ x^s(t;0) + x^s(t;2b)]/2
 + {a\over 2b} \nonumber  \\
&& \times
 \begin{cases}
{1\over 1-s\nu} [x^{s-\mu}(t; 2b)-
  x^{s-\mu} (t; 0)], & $if $\mu \ne s \cr
  s\log [x(t;0)/
 x(t;2b)],  & $if $\mu =s \cr  \end{cases}
\end{eqnarray}
where
\begin{equation}
x(t; b) = \bigl[x_0^{-\mu} e^{-\mu at} + b(1-e^{-\mu at})/a
 \bigr]^{-1/\mu}
\end{equation}
is the solution of Eq.~(2) for a constant $a$ and $b$ parameters.
Using the asymptotic forms
$x(t; 0) = x_0 e^{at}$ and  $x(t; 2b) \to  (a/2b)^\nu$,
one concludes that the  higher order, $s >\mu$,  transient
moments diverge exponentially
\begin{equation}
\langle x_t^s \rangle  \propto e^{(s-\mu)at}
\end{equation}
and for the marginal case $s=\mu$ $(=\lambda/a)$
\begin{equation}
\langle x_t^\mu \rangle  \to  {\rm const} + {a^2 \mu \over 2b} t
\end{equation}
grows linearly for long times. Note that according to (15) or (9)
the value of the ratio $\alpha=\lambda/a$
(frequency of switching by the rate of a free growth)
specifies the order of the lowest divergent moment. In the
``double--critical'' case of exact Pareto distribution (17)
it is simultaneously
equal to the value of a ``degree of nonlinearity'' $\mu$.
Thus, for the most important true Verhulst
or so--called Stratonovich (with $\mu=2$,
corresponding to the quartic potential) model it means that
it is the mean value or the variance, respectively,
that grows linearly with time, Eq.~(21).

\section{Conclusions}
The anomalous systems exhibit large fluctuations which are
difficult to explain dynamically within standard theories,
for instance, as a result of an additive
Gaussian diffusion.
Thus such a behavior is frequently joined with the presence
of a ``large'' driving noise, e.g., a Levy noise or a
multiplicative Gaussian noise. On the other hand it was
suggested that certain aspects of such behavior are rather
related to some hierarchical structure of kinetic processes
at the deterministic level [2,3,5]. This idea was exploited
in the present work for exactly solvable generic model
(2), (5). The general behavior of the system
depends on two things, on the mutual relation between
$\lambda$ and $a$ (or $\mu a$) and on the value of $\sigma^2$,
namely whether $\sigma^2 <1$ or $\sigma^2 =1$.

Let $\sigma^2 <1$  (i.e., below the critical line in Fig.~1).
If frequency $\lambda$ is smaller than
the rate of relaxation processes $\mu a$  the $x_t$
``hardly'' follows $b(t)$, Eq.~(5), switching between some
values located close to $x_l$ or $x_r$, Eq.~(14),
respectively (right part of plot).
Successive  (``rare'' and  ``short'')
moves up and down are associated with
values $b(1 - |\sigma|)$ and $b(1 + |\sigma|)$ in Eq.~(2),
respectively, and in principle look like different
deterministic processes. In the opposite case
(left part of plot), $\lambda > \mu a$,
the evolution looks more like the Verhulst kinetics with
the averaged value $b = \overline{b(t)}$ of annihilation coefficient.
The distribution of $x$ is unimodal, however still with a rather wide
maximum, located somewhere in a middle of the support.
The value $\lambda =\mu a$ is critical for so--called noise
induced transition. This transition is formally controlled by
correlation--time of the noise.

The case of $\sigma^2 =1$ is of particular interest for
the present work and may be formally considered
as the ``true''
transition, which is indicated by singular
dependence of moments and related power--laws.
This transition is controlled by the noise intensity.
The asymptotic distribution of $x$ is given by Eq.~(15).
$P(x)$ has a fat--tail with the index $\alpha= \lambda/a$.
The distributions with $\alpha <2$ belong --- via central limit
theorem --- to the basin of attraction of appropriate Levy
$\alpha$--stable distribution. Thus the  kinetic (2), (5)
with randomly flashing annihilation term may be treated as
a source  of a Levy noise at the macroscopic level, if
the frequency of switching is small enough, $\lambda < 2a$, or,
equivalently, if the correlation--time is sufficiently
long $\tau > (4a)^{-1}$. For the Stratonovich model, which
is presented in Fig.~(1), this corresponds precisely to
positions on the line right from the critical point
[with $\tau = (2\mu a)^{-1}$]. If the frequency  is large,
$\lambda >2a$, or the correlations are short,
$P(x)$ still has a fat--tail, however it corresponds to the
Gaussian distribution via central limit theorem.


\begin{figure}
\caption{The phase diagram on $(\tau,\, D)$--plane, Eq.~(16),
for Stratonovich model $\mu =2$ and dimensionless units $a=b=1$.
Plots of normalized $P(x)$  vs $x$ for particular values
 of $(\delta, \sigma) \in \{1/4,\,1/2,\,2\}\times \{3/10,\,1\}$.
}
\end{figure}


\begin{thebibliography}{}
\bibitem{1}
M.  B. Weissman, Rev. Mod. Phys. {\bf 60}, 537 (1988).
\bibitem{2}
S. M. Bezrukov, in {\it Proceedings of the First International
 Conference on Unsolved Problems of Noise,
Szeged, 1996}, edited by C. R. Doering, L. B. Kiss, and M. F.
 Schlesinger
(World Scientific, Singapore, 1997), pp. 263--274;
S. M. Bezrukov and M. Winterhalter, Phys. Rev. Lett. {\bf 85}, 202
 (2000).


\bibitem{3}
J. Davidsen and H. G. Schuster,
 Phys. Rev. E {\bf 62}, 6111 (2000).



 \bibitem{4}
 T. Antal, M. Droz, G. Gy\"orgyi, and Z. R\'acz,
 Phys. Rev. Lett. {\bf 87}, 240601 (2001).
\bibitem{5}
 Z. Siwy and A. Fuli\'nski,
 Phys. Rev. Lett. {\bf 89}, 158101 (2002).
\bibitem{6}
B. Kaulakys and J. Ruseckas,
 Phys. Rev. E {\bf 70}, 020101 (2004).

\bibitem{7}
J.-P. Bouchaud and A. George, Phys. Rep. {\bf 195}, 127 (1990).

\bibitem{8}
O. Biham, O. Malcai, M. Levy, and S. Solomon, Phys. Rev. E {\bf
 58}, 1352 (1998);
O. Malcai, O. Biham, P. Richmond, and S. Solomon,
 Phys. Rev. E {\bf 66}, 031102 (2002).

\bibitem{9}
Z. Burda, D. Johnston, J. Jurkiewicz, M. Kami\'nski, M. A. Nowak, G.
 Papp, and I. Zahed,
Phys. Rev. E {\bf 65}, 026102 (2002).

\bibitem{10}
 R. Metzler and J. Klafter, Phys. Rep. {\bf 339}, 1 (2000)


\bibitem{11}
S. Jespersen, R. Metzler, and H. C. Fogedby,
 Phys. Rev.  E {\bf 59}, 2736 (1999);
I. M. Sokolov, Phys. Rev. E {\bf 63}, 011104 (2001);
{\bf 63}, 056111 (2001);
I. M. Sokolov and R. Metzler,
Phys. Rev. E {\bf 67}, 010101(R) (2003).

\bibitem{12}
 B. Dybiec and E. Gudowska-Nowak,
 Phys. Rev. E {\bf 69}, 016105 (2004).



\bibitem{13}
W.~Horsthemke and R.~Lefever,
{\em Noise-Induced Transitions} (Springer, Berlin, 1984);
 C.R.~Doering
and W.~Horsthemke, J.\ Stat.\ Phys.\ {\bf 38}, 763 (1985);
A.~Teubel, U.~Behn,
and
A. Kuehnel, Z. Phys B {\bf 71}, 393 (1988);
A. Fuli\'nski, Phys.\
Rev.\ E {\bf 52}, 4523 (1995); Acta Phys.\ Pol.\ B
{\bf 27}, 767 (1996).

 \bibitem{14}
 R.~Zygad\l{}o, Phys.\ Rev.\ E {\bf 54}, 5964 (1996).


\bibitem{15}
I.L.~Luke, {\em The Special Functions and Their Approximations}
(Academic, New York, 1969); I.S.~Gradshteyn, I.M.~Ryzhik, {\em
Table of Integrals, Series and Products} (Academic,
New York, 1965).



\end{thebibliography}
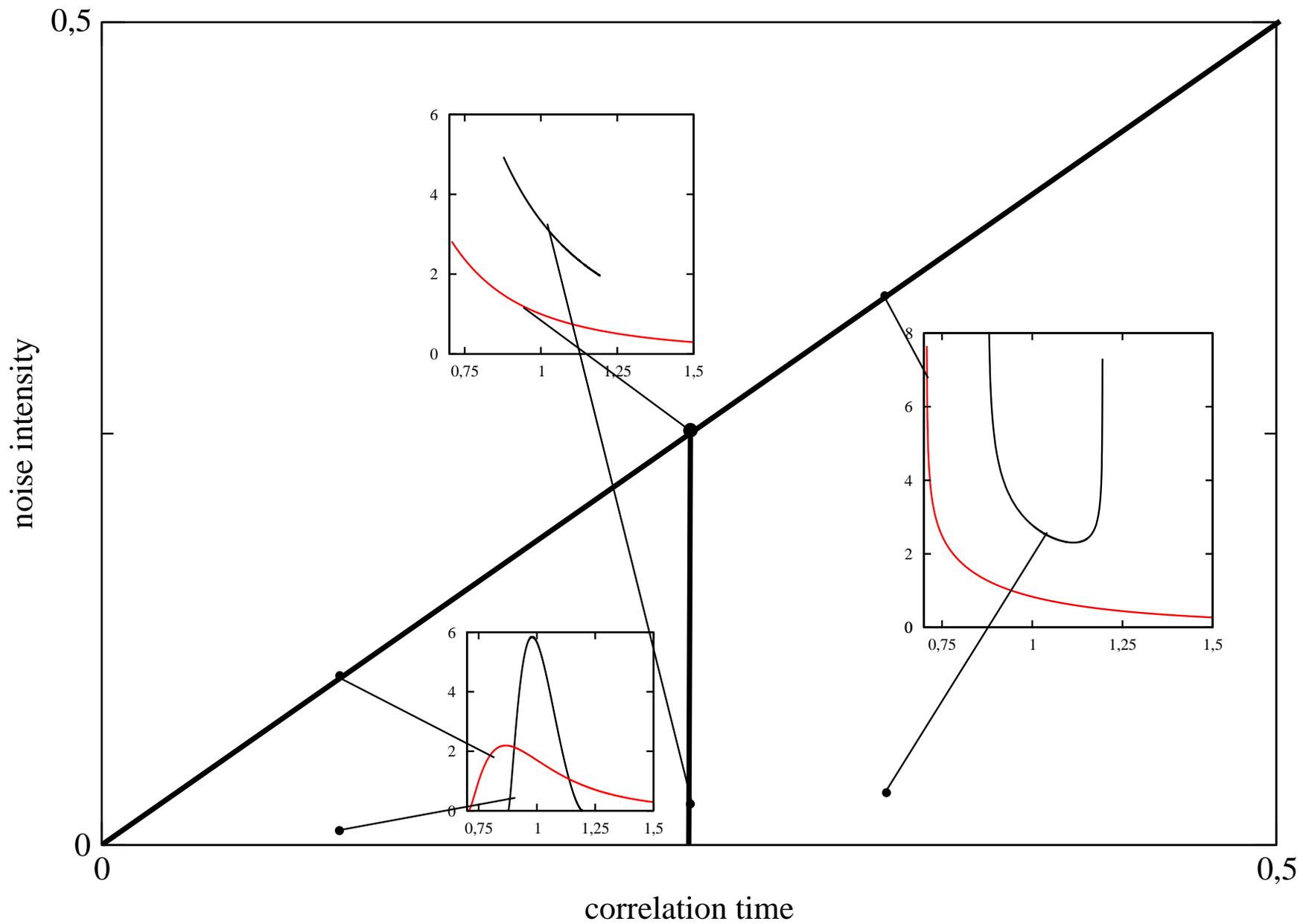
\end{document}